\documentclass[]{emulateapj}
\usepackage{amsmath}
\usepackage{natbib}
\slugcomment{Draft version}

\shorttitle{Gravitational lens modeling with basis sets}
\shortauthors{Birrer et al.}

\begin{document}

\title{Gravitational lens modeling with basis sets}

\author{Simon Birrer}
\email{simon.birrer@phys.ethz.ch}
\author{Adam Amara}
\email{adam.amara@phys.ethz.ch}
\author{Alexandre Refregier}
\email{alexandre.refregier@phys.ethz.ch}

\affiliation{Institute for Astronomy, Department of Physics, ETH Zurich, Wolfgang-Pauli-Strasse 27, 8093, Zurich, Switzerland}

\date{\today}

\begin{abstract}

We present a strong lensing modeling technique based on versatile basis sets for the lens and source planes. Our method uses high performance Monte Carlo algorithms, allows for an adaptive build up of complexity and bridges the gap between parametric and pixel based reconstruction methods. We apply our method to a  \textit{HST} image of the strong lens system RXJ1131-1231 and show that our method finds a reliable solution and is able to detect substructure in the lens and source planes simultaneously. Using mock data we show that our method is sensitive to sub-clumps with masses four orders of magnitude smaller than the main lens, which corresponds to about $10^8 M_{\odot}$, without prior knowledge on the position and mass of the sub-clump.  The modeling approach is flexible and maximises automation to facilitate the analysis of the large number of strong lensing systems expected in upcoming wide field surveys. The resulting search for dark sub-clumps in these systems, without mass-to-light priors, offers promise for probing physics beyond the standard model in the dark matter sector.

\end{abstract}


\maketitle

\section{Introduction}
The standard cosmological model is based on the standard model of particle physics, Einsteins theory of General Relativity, a cosmological constant, cold dark matter and inflation. The physical origin of the cosmological constant, inflation and dark matter remains a mystery to date. The predictions of the expansion history of the universe has been probed with high precision and structure formation has been tested from the horizon scale down to about 1Mpc or even below \citep[e.g.,][]{Collaboration:2015p9875, Dawson:2013p9894}. The smallest scale tests come from the Lyman-alpha forest \citep[see e.g.,][]{Seljak:2006p9712} and strong and weak lensing in anomalous quadrupole lenses \citep[e.g.,][]{Inoue:2014p9963}. At even smaller scales in the non-linear regime, there are observational and theoretical challenges in bringing model and data in agreement. This problem occurs predominantly in the number, phase space densities and density profiles when comparing simulations of dark matter substructure with observations of luminous satellite galaxies in our Milky Way \citep[see e.g.,][]{Kauffmann:1993p5842, Klypin:1999p4649, Moore:1999p4657, BoylanKolchin:2011p4244}. A potential non-gravitational (i.e. collisional) effect of a dark matter particle may have an effect on structure formation on small scales without having an effect on larger scales. Probing the small scale structure formation and mass distribution may thus provide information beyond the $\Lambda$CDM model. 

Strong lensing is a powerful probe to test structure formation on small scales \citep[][]{Metcalf:2001p9744, Dalal:2002p10031, Yoo:2006p10063, Keeton:2009p9737, Moustakas:2009p9694}. Strong lensing is the effect caused by the bending of light by massive foreground over-density (e.g. galaxy, group or cluster) such that multiple images of the same background object appears. This effect is well suited for many astrophysical and cosmological applications \citep[see e.g.][a review focused on galaxy sized lenses]{Treu:2010p4708}. Strong Lensing was also proposed to detect luminous and dark substructure in the lens \citep[][]{Koopmans:2005p8841, Vegetti:2009p9255}. This technique has been successfully applied to data \citep[][]{Vegetti:2010p9515, Vegetti:2012p4937} where sub-clumps down to about $2 \times 10^8 M_{\odot}$ masses have been detected. Substructure also has an effect on the flux ratios in multiple lensed quasar images \citep[see e.g.,][]{Metcalf:2002p9547, Kochanek:2004p10033, Amara:2006p4715, Metcalf:2012p5467, Xu:2015p9516}. Anomalous flux ratios have thus been reported in the literature. \cite{Takahashi:2014p9660} pointed out that the anomalous flux ratios measured can be accounted by line-of-sight structure and do not have to necessarily come from structure within the lens. With recent and upcoming large scale surveys new area and depth becomes available to discover strong lens systems. \cite{Oguri:2010p10017} forecasted thousands of lensed quasar systems from DES and LSST. These datasets will help to constrain the statistical features of the small scale structure imprinted in the strong lensing signal. The increasing number of strong lens systems will in the future need to be analyzed with automated modeling approaches.

The aim of this paper is to describe a lens modeling approach that can be applied to different lens systems without adjusting parameter priors by hand and uses all the information contained in a image to constrain the projected mass density of the lens with a special emphasis on substructure. Our model approach is based on parameterized basis sets in the source surface brightness and lens model. The model framework can handle an adaptive complexity in the source and lens models. In addition to the basis sets, we show the power of modern sampling techniques and we make use of fast  computational methods. 

The paper is structured as follow: In Section \ref{sec:overview} we give an overview of existing lens model techniques and show how they relate to our modeling approach. In Section \ref{sec:basis_sets} the source surface brightness and lens potential basis sets on which our model relies on are introduced. Section \ref{sec:fitting} describes the model fitting procedure and in particular how the source surface brightness reconstruction is done and how we deal with the high number of non-linear parameters in the lens model. We test our fitting procedure on mock data and on Hubble data of the lens system RXJ1131-1231. In Section \ref{sec:detection}, we study how well we can detect substructure in a lens model without prior information on the mass, slope and position. This section is followed by a conclusion (Section \ref{sec:conclusion}).

\section{Overview of Lens Model techniques} \label{sec:overview}
Galaxy-size strong lenses have been modeled extensively in the literature (see references below in this section). 
The following aspects have to be modeled in a strong lens system when comparing data and model on the image level:
\begin{itemize}
	\item the lens mass model
	\item the source surface brightness profile
	\item the lens surface brightness profile
	\item the point spread function (PSF)
\end{itemize}
Depending on the lens system and instrument, one has to also model dust extinction, external convergence, micro lensing by stars and other aspects.

Depending on the scientific aim, the main focus is typically more on the source surface brightness reconstruction or on the lens mass reconstruction. In both cases one can, broadly speaking, divide the modeling techniques in two regimes:

(1) Parametric reconstruction: Using simple and physically motivated functional forms with a controllable number of parameters ($\sim$ 10) \citep[e.g.,][]{Kochanek:1991p8926, Kneib:1996p8941, Keeton:2001p8965, Jullo:2007p9020}. A controllable number of parameters implies that one can fully explore the parameter space and convergence to the best fitting configuration can often be obtained.

(2) Pixel based reconstruction: This is most often done with a grid where each pixel is treated as a free parameter. Pixelised source surface brightness inversions have been proposed by e.g. \cite{Wallington:1996p8883, Warren:2003p8842, Treu:2004p8886, Dye:2005p8890, Koopmans:2005p8841, Brewer:2006p8800, Suyu:2006p5328, Wayth:2006p8894, Suyu:2010p4938}. These methods often rely on a regularization of the pixel grid when there is not a unique solution. Depending on the regularization procedures, priors and the pixel size, one can come to different reconstructed sources \citep[see e.g.][]{Suyu:2006p5328, Suyu:2013p4952}. Recently \cite{Tagore:2014p9985} did an analysis of statistical and systematic uncertainties in pixel-based source reconstructions. Furthermore, these methods are computationally expensive as they rely on large matrix inversions. For the lens mass or its potential, grid based modeling has been applied by e.g. \cite{Blandford:2001p9413, Saha:2004p9395, Bradac:2005p9147, Koopmans:2005p8841, Saha:2006p9215, Suyu:2006p9219, Jee:2007p9229, Vegetti:2009p9255, Suyu:2009p8893, Vegetti:2012p4937, Coles:2014p8364} and even mesh-free models \cite{Merten:2014p9999}.

Computational techniques also vary for different modeling approaches. Ray-tracing has generally been used to map extended surface brightness from the source to the image plane. If significant surface brightness variations occur on very small scales, such as for quasars due to their compact size, simple ray-tracing can lead to numerical inaccuracies. One way to model such systems is to approximate quasars as point sources. One then solves the lens equation numerically for the positions in the image plane \citep[recently e.g.][]{Suyu:2013p4952}. An alternative to avoid the point source approximation is adaptive mesh refinement \citep[e.g.,][]{Metcalf:2012p5467, Metcalf:2014p9779} which changes the ray-tracing refinement scale depending on the local spacial variation of the source at different image positions.

In standard $\Lambda$CDM, the self-similarity of dark matter indicates that the same amount of complexity as seen in galaxy clusters must also be present in galaxy-sized strong lens systems. Its effect is much weaker in terms of deflection and magnification, but it must still be present. Ideally, we want a model that is flexible such that it can describe any lens mass and source surface brightness distribution. For this model we need to be able to explore its degeneracies and to converge to the `true' solution to extract the information contained in a strong lens system.

One of the aims of our work is to fill the gap between the parametric and non-parametric models. We do so by choosing basis sets that we treat in a fully parametrized form.

\section{Choice of basis sets} \label{sec:basis_sets}
In the following sections we describe our choices for basis sets and, in addition, we present how we produce mock data given a set of parameter values.

\subsection{Basis for the source} \label{sec:basis_source}
We make use of shapelets \citep[introduced by][]{Refregier:2003p8153, Refregier:2003p8257, Massey:2005p10064} in the source surface brightness plane. We implemented the two-dimensional Cartesian shapelets \citep[Eq. 1 and 18 in][or in Appendix \ref{app:shapelets} of this work]{Refregier:2003p8153}. Independent of this work \cite{Tagore:2015p10068} proposed a different method to use shapelets in the source reconstruction. Shapelets form a complete orthonormal basis for an infinite series. Restricting the shapelet basis to order $n$ provides us with a finite basis set that is linked to the scales being modeled. If we wish to model a larger range of spatial scales in the surface brightness profile, we need to use more high order shapelets. The number of basis functions $m$ is related to the restricted order $n$ by $m=(n+1)(n+2)/2$. The shapelet basis functions allow us to dynamically adapt to a given problem. We can increase the complexity when we need them and reduce it when it is not appropriate. Apart from the order $n$ one can also set the reference scale $\beta$ of the basis function. Minimal and maximal scales ($l_{\text{min}}$, $l_{\text{max}}$) being resolved up to order $n$ with reference scale $\beta$ is $l_{\text{min}} = \beta/\sqrt{n+1}$ and $l_{\text{max}} = \beta\sqrt{n+1}$. The parameter $\beta$ is a user specified choice. Another choice is the peak position of the shapelet center $(x_0, y_0)$. For any finite order in $n$, the choice of the center is crucial for the fitting result. A natural choice for $(x_0, y_0)$ is the center of the light profile of the source galaxy. In that sense $(x_0, y_0)$ must be interpreted as two non-linear parameters. 

\subsection{Basis for the lens} \label{sec:lens_model}
Choosing a realistic basis set for the lens mass distribution is a challenging task as there are many different scales involved, especially when considering low mass sub-clumps. These sub-clumps are very small in scale but are also very dense. Having a basis set which allows a general description of such clumpy halos on different scales typically involves a large number of parameters. Depending on the sub-clump mass limit being considered, there are hundreds or even thousands of sub-clumps expected. A minimal description requires at least information about individual positions, masses and concentrations. Such a description leads to a degenerate and non-unique lens model \citep[e.g.][]{Keeton:2010p8371, Kneib:2011p8381}. For cluster lenses, the typical masses of substructure are several orders of magnitude below the total lens mass, but it is possible to give strong priors on the location of the substructure, namely at the position of the luminous galaxies. For detecting invisible substructure such a prior can not be used. As often called `non-parametric' or `free-form' approach, meaning there are more parameters than data constraints (i.e. deliberately under-constrained) was proposed and implemented by \cite{Saha:2004p8382} and \cite{Coles:2014p8364}. Using the catalog level image position information and time-delay measurements, there is far less information available than parameters to be constrained. One is able to draw random realizations of lens models that meet all the constraints. Statements about the validity of a specific lens model can only be drawn statistically. Doing a comparison on the image level where about $10^3$ - $10^4$ pixels are involved, more information is available to constrain the model.

In our approach we start with a softened power-law elliptical potential (SPEP) \citep[e.g., discussed by][]{Barkana:1998p5324}. The lensing potential $\Phi$ is parameterized as
\begin{equation} \label{eqn:SPEP}
 \Phi(x_1, x_2) = \frac{2E^2_p}{\eta^2} \left( \frac{\rho^2_p + s_p^2}{E^2_p} \right)^{\eta/2}
\end{equation}
where
\begin{equation}
 \rho^2_p = x_1^2 + x_2^2/\cos^2\beta_p
\end{equation}
with $\cos \beta_p$ being the axis ratio of the potential, $\eta$ the power-law index, $E_p$ the normalization of the potential, $s_p$ the smoothing length and $x_{1,2}$ the position rotated such that $x_1$ is in the direction of the major axis of the potential. For an additional sub-clump, we model them either as a spherical NFW \cite{Navarro:1997p8389} profile or a spherical power-law potential (SPP). For both functions, we set the softening length $s_p = \text{const} = 0.0001"$ for computational reasons. In that sense the softening is virtually zero and is not a free parameter in this work.

Combining the two functions (SPEP and SPP) we get $6+4=10$ non-linear free and partially degenerated parameters to be fitted. With this parameterization we expect a good overall fit to many different lens systems and perhaps to catch the largest substructure within the lens, visible or invisible. Such tests are shown in section \ref{sec:example}.

In addition, we include two dimensional Cartesian shapelets (same functional form as for the source in Section \ref{sec:basis_source}) in the potential. We choose the scale factor $\beta$ to be the Einstein radius. This allows for perturbations at the global scale of the lens that can not be made with another peaked profile. The first derivatives of the potential (deflection angle) and second order derivatives (convergence and shear) can be computed analytically and can be expressed within the same shapelet basis functions (See Appendix \ref{app:shapelets}), thus enabling fast computations.

\subsection{Basis for the lens light}
For the description of the lens light, we use S\'ersic profiles \citep[][]{Sersic:1968p10108}. Depending on the lens galaxy, adding multiple S\'ersic profiles can lead to better fits \citep[see e.g.][]{Suyu:2013p4952}.

\subsection{Image making} \label{sec:image_making}
Having a parametric description of the source surface brightness, a possible point source, the lens potential, the lens light and the PSF, an image can be generated in the following steps:
\begin{enumerate}
\item Starting in the image plane one evaluates the analytic expression of the deflection angle using grid based ray-tracing. The resolution has to be of order (or slightly smaller than) $\beta/\sqrt{n}$ to capture the features in the extended source model. 
\item We then compute the point source image in a iterative ray-shooting procedure starting from the local minimas of the relative distance to the point source of step 1. Corrections for the next proposed ray-shooting position can be made when considering the relative displacement to the point source and the second order derivatives of the lens potential. The requirement of the precision of the point source position in the image plane of about 1/1000 of the pixel size can be reached within very few iterations.
\item For the point sources, which appear as PSF's, we normalize the externally estimated PSF to their intrinsic brightness and lens magnification. We do not lose significant computational speed when modeling the PSF further out to the diffraction spikes. For the extended surface brightness a numerical convolution needs to be made. This can be done either at the pixel or sub-pixel level. This step is the most expensive computational process in the forward image modeling. The process scales roughly linearly with the number of pixels or sub-pixels in the convolution kernel. We use Fast-Fourier-Transforms implemented in a {\tt scipy} routine in {\tt python}. Our default kernel size is $15 \times 15$ pixels.
\item The lens light is added with analytical Sersic profiles convolved with the same PSF kernel as the extended source surface brightness.
\end{enumerate}
For the modeling, we do not add noise. When simulating realistic images, we add a Gaussian background noise with mean zero to all pixels and a scaled Poisson noise on the signal (pixel by pixel).

\section{Model fitting} \label{sec:fitting}
For the modeling, we have three questions to answer:
\begin{enumerate}
	\item What is the best fit configuration of the model to match the data of a specific lens system? We want to find the global minima for the $\chi^2$ value. \label{Q1_1}
	\item What level of complexity is needed to fit the data to a certain level? We want to compare consistency with the data by analysing the reduced $\chi^2$ value and compare different model configurations with a Bayes factor analysis. \label{Q2_1}
	\item How well is the model solution determined by the data? We want to sample the parameter space and determine confidence intervals. \label{Q3_1}
\end{enumerate}
As a result, many choices have to be made in the lens modeling. More than 10 parameters in the lens model with non-linear behavior have to be specified. For a realistic surface brightness description the shapelet order $n$ can be higher than 20 which corresponds to 154 basis' and their corresponding coefficients. Given this level of complexity, even the first question on its own is difficult to address. Once we have a method for addressing the first question, repeating the procedure with different choices of complexity and parameterization will provide an answer to question \ref{Q2_1}. Question \ref{Q3_1} can then be answered with a Bayesian inference method such as a Markov chain Monte Carlo (MCMC) sampling. For this we use the software package \texttt{CosmoHammer} \citep[][]{Akeret:2013p8317}, which is based on the \textit{emcee} method of \cite{Goodman:2010p8363} and its implementation by \cite{ForemanMackey:2013p8357}. The software package allows for massive parallelization in the sampling process. In this section, we focus on question \ref{Q1_1}. We will describe in detail the methods and procedures we apply to make the algorithm converge to the best fit lens model configuration. Question \ref{Q2_1} and \ref{Q3_1} are addressed with examples in Section \ref{sec:example} and \ref{sec:detection}.

\subsection{Source surface brightness reconstruction} \label{sec:reconstruction}
In our method we use a weighted linear least square approach to reconstruct the source surface brightness. This is a standard procedure to minimize the quadratic distance between data and model with weighted error measures. The estimation of the covariance can also be calculated (see Eqn \ref{eqn:min_problem} - \ref{eqn:cov_matrix} below). The minimization problem has to be linear. Let $\vec{y}$ be the data vector of dimension $d$. In our system, it contains all the pixel values of the image in the area of interest for a surface brightness reconstruction. Let $W$ be the weight matrix of dimension $d \times d$. In a likelihood interpretation, $W$ is the inverse covariance matrix of the data. Assuming the pixel errors are uncorrelated $W$ is a diagonal matrix. Let $\vec{\xi}$ be the parameter vector of dimension $m$. In our case, $\vec{\xi}$ is the vector of the coefficients of the linear combination of shapelet basis functions. The number of shapelet basis functions $m$ depends on the shapelet order $n$ as described in section \ref{sec:basis_source}. Let $X$ be the linear response matrix of the shapelet parameters on the pixel values in the image plane of dimension $d \times m$. The product $X\vec{\xi}$ describes a lensed and convolved surface brightness on the image plane. $X$ can be computed by mapping all $m$ shapelet basis functions from the source to the image plane, convolve and resize them separately on the pixel scale. The computational cost of this procedure is linear in the number of basis functions involved and dominates the process for low $m$.

\begin{figure}
  \centering
  \includegraphics[angle=0, width=88mm]{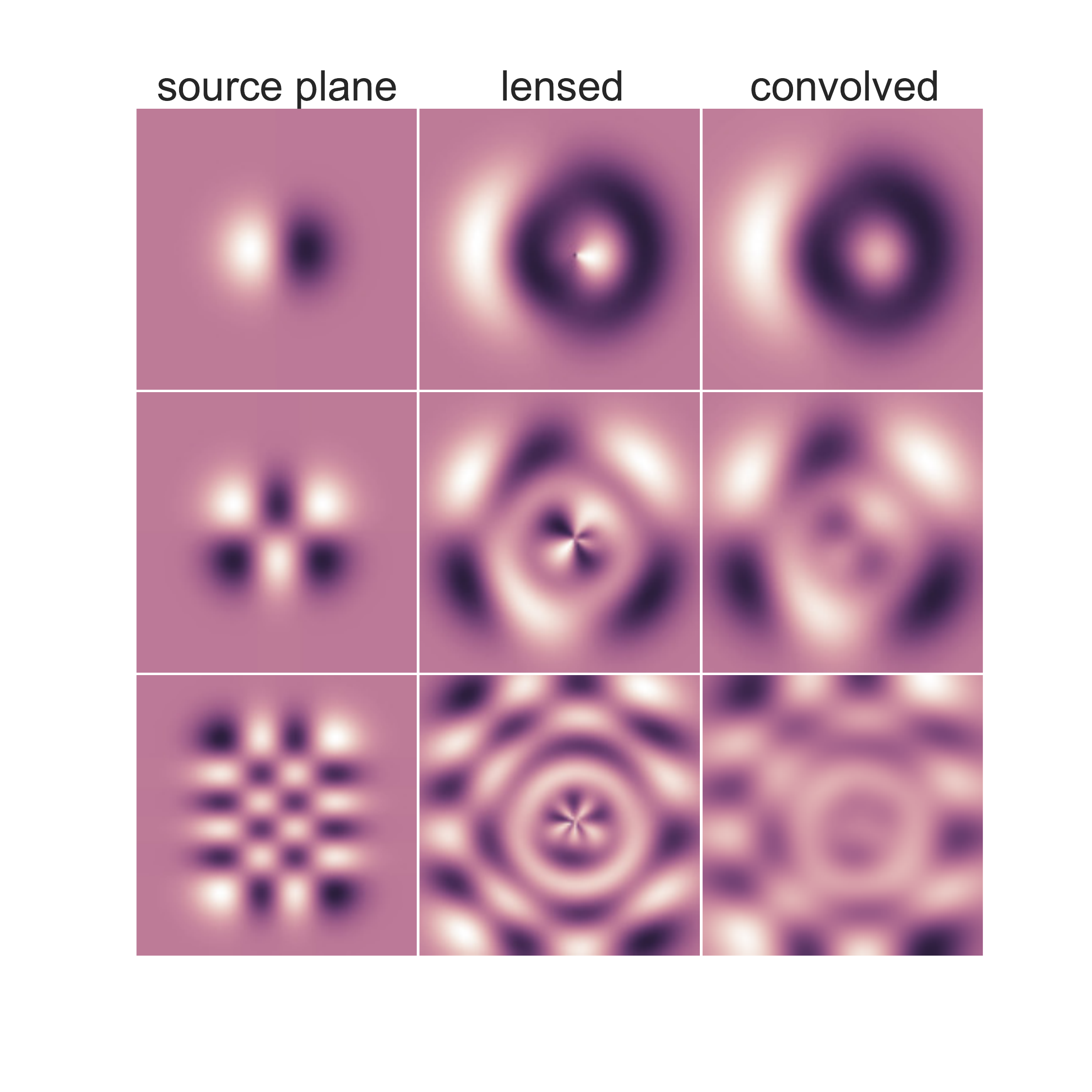}
  \caption{An illustration of the modeling of the source surface brightness with three different shapelet basis functions. Left panels: Shapelet basis function in the source plane. Middle panels: Mapped shapelets in the image plane with a SIS lens via ray-tracing. Right panels: PSF convolved image. From top to bottom: Shapelets with $(n_1,n_2)=(1,0), (2,1), (3,5)$.
  }
\label{fig:shapelet_lens}
\end{figure}

Figure \ref{fig:shapelet_lens} illustrates how the shapelet basis functions are mapped. The problem of finding the best source configuration $\vec{\xi}_0$ given the data $\vec{y}$ and the weights $W$ can be posed as a weighted linear least square problem:
\begin{equation} \label{eqn:min_problem}
	\vec{\xi}_0 = \text{min}_{\xi} \|W^{1/2}(\vec{y}-X\vec{\xi})\| 
\end{equation}
This equation can be written as
\begin{equation}
	(X^\top WX)\vec{\xi}_0 = X^\top W \vec{y}
\end{equation}
whose solution is given by
\begin{equation} \label{eqn:beta_max}
	\vec{\xi}_0 = (X^\top WX)^{-1}X^\top W \vec{y}.
\end{equation}
The covariance matrix of $\vec{\xi}$, $M^{\xi}$ is therefore given by
\begin{equation} \label{eqn:cov_matrix}
	M^{\xi} = (X^\top WX)^{-1}.
\end{equation}
$M^{\xi}$ becomes important when marginalizing the probability distribution over $\vec{\xi}$.

The procedure involves a matrix inversion of dimension $m \times m$. The computational cost and memory allocation of this inversion becomes more significant with larger $m$. Moreover, the matrix $(X^\top WX)$ has to be invertible. If not, this method fails to find a solution and regularization is needed.
A grid based regularization was introduced by \cite{Warren:2003p8842}. Conceptually and computationally, the method of \cite{Warren:2003p8842} and the one presented in this paper differ significantly. The matrix $(X^\top WX)$ is a dense matrix where as the matrix in grid based regularization can be sparse. A sparse matrix can only be maintained when having a small PSF(e.g. $5\times5$ pixel). We use in our method a default PSF kernel of $15\times15$ pixels and a further extension affects only the FFT-convolution of the lensed shapelet basis functions. Our method is well suited to reconstruct also lensed sources in images with larger PSF's than HST images. But the main gain of our method is in terms of the number of parameters (i.e. the size of the matrix). Well chosen basis sets can allow for a smaller number of parameters compared to grid based methods significantly.

\begin{figure*}
  \centering
  \includegraphics[angle=0, width=180mm]{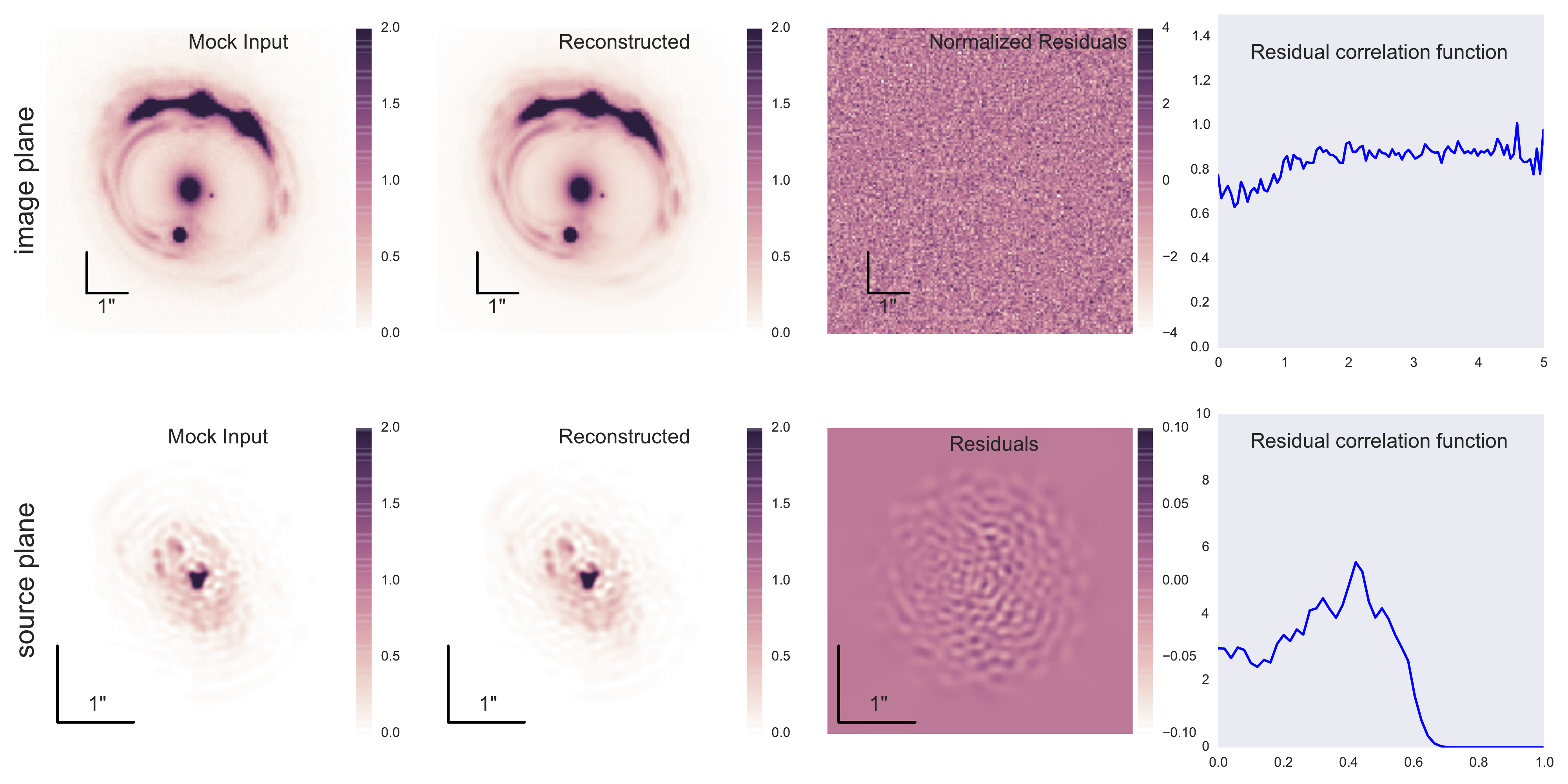}
  \caption{Demonstration of the source surface brightness reconstruction with upper panels showing the image plane and lower ones for the source plane. From left to right: Initial mock image (source), reconstructed image (source), relative residuals, 1D correlation function of residuals. The image is almost perfectly reproduced even without significant residual correlations. The features of the source surface brightness profile is very well reproduced. The relative intensities of input vs. output is 10\% or below. The spacial correlation of the relative difference is enhanced. This feature reflects the properties of the shapelet basis functions involved and the minimal and maximal scales of those.}
\label{fig:reconstruction}
\end{figure*}

In Figure \ref{fig:reconstruction}, we take a mock image produced with a chosen $\vec{\xi}_0$ (incl. point sources) with maximum shapelet order $n_{\text{max}}=40$ and added Poisson and Gaussian background noise to the image. We check the reconstruction by computing the relative residuals and their correlation function. We see from Figure \ref{fig:reconstruction} that the results are almost consistent with purely uncorrelated noise. Only for very small separations, the correlation is marginally smaller than with noise. This effect highly depends on the signal-to-noise ratio of the shapelets. Since we know the input source surface brightness, we can also check its reconstruction. The error in input vs. output in the source has features which represent the scales of the shapelet basis functions. The relative error of the surface brightness is about 10\% or less. This reconstruction process with $n_{\max} = 40$ and $15\times15$ pixel convolution kernel takes about 4s on a standard personal computer. When reducing the number of shapelet coefficients to $n_{\max} = 20$ the reconstruction falls below 1s.

\begin{figure}
  \centering
  \includegraphics[angle=0, width=88mm]{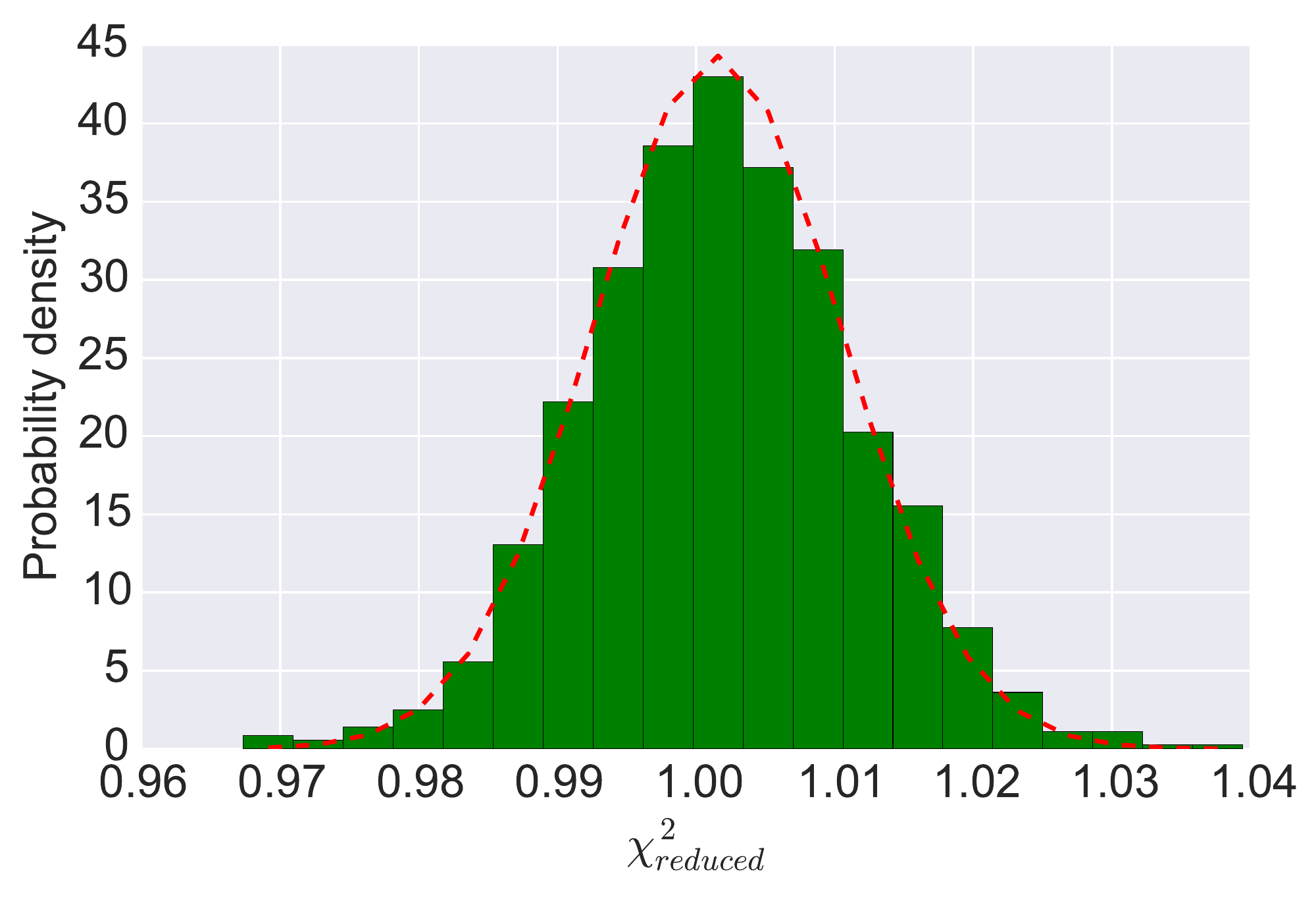}
  \caption{The distribution of the reduced $\chi^2$ values of 1000 realizations of the image reconstruction process (see Figure \ref{fig:reconstruction}) is plotted. Each realization differ in the Gaussian and Poisson noise realization. The mean value of the distribution is $\chi^2_{\text{red}} = 1.0015$ and the spread is $\sigma = 0.009$.}
\label{fig:chi2_distribution}
\end{figure}

The specific reconstruction depends on the noise realization. By repeating the reconstruction 1000 times with different noise realizations, we find that the reconstruction is stable. In Figure \ref{fig:chi2_distribution} we plot the reduced $\chi^2$ distribution of the different realizations. We find a mean $\chi^2_{\text{red}} = 1.0015$ with a standard deviation of $\sigma = 0.009$.

\subsection{Convergence techniques} \label{sec:convergence}
In the previous section, we showed that we can linearize all parameters of the source model given a specific lens model and thus we can express it as a linear minimization problem. The marginalization of the linear parameters can be made analytically (see Section \ref{sec:likelihood_compute} below). Changes in the lens model however have a non-linear effect on the image. In that sense, we can marginalize over many parameters in our model and are left with about 10-30 non-linear parameters. To explore this space we use a Particle Swarm Optimization (PSO) \cite{Kennedy:2001p8447} algorithm to find the minimum of the parameter space. The algorithm is described in more detail with an illustration in Appendix \ref{app:pso}.

Convergence towards the global minimum in parameter space can depend on several factors. It depends on the volume of the parameter space, the number of local minima and the shape of the cost function around the absolute minimum. As we are marginalizing over all the source surface brightness parameters, one can have unexpected behavior of the cost function over the lens parameters. In the following we describe our convergence method which goes beyond simply applying the PSO algorithm and which is important for the performance of our method.

\begin{figure*}
  \centering
  \includegraphics[angle=0, width=180mm]{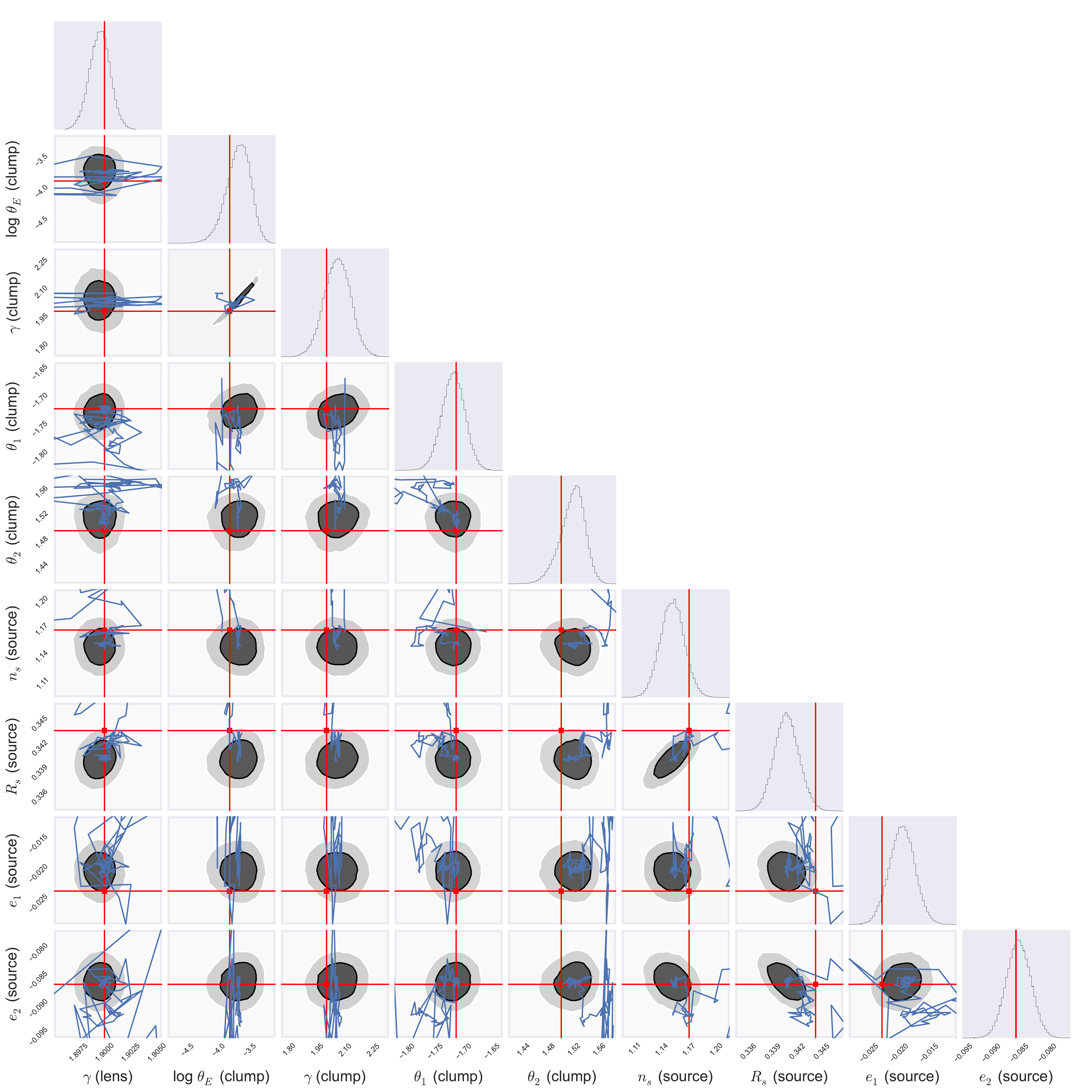}
  \caption{Illustration of a combined PSO and MCMC chain in a 9 dimensional non-linear parameter space. The blue lines connect the best fit particle during the PSO process. The red lines mark the true input parameter. Dark (light) gray contours mark the 68\%-CL (95\%-CL) interval estimated from the MCMC process.}
\label{fig:mock_triangle}
\end{figure*}

\subsubsection{Parameterization}
The sampling in parameter space can be made in any parameterization with a bijective transformation to the originally described form. The parameterization can have a significant impact on the convergence capacity and performance of a specific algorithm. If there are periodic boundaries in a specific parameterization, some algorithms can have difficulties. In our model, this is the case for the parameter of the semi-major axis angle of the elliptical lens potential $\theta_0$ which is defined in the range $[0,\pi)$. The model can continuously rotate the axis but the parameter space has to jump from 0 to $\pi$, or vise versa. Mapping $\theta_0$ and the axis ratio $ q = \cos \beta$ into ellipticity parameters $(e_1, e_2)$ with $f : [0,\pi) \times (0,1] \rightarrow (-1,1) \times (-1,1)$ given as
\begin{equation}
	f\left( \theta, q \right) = \left( \frac{1 - q}{1 + q} \cos(2\theta) , \frac{1 - q}{1 + q} \sin(2\theta) \right)
\end{equation}
provides a continuous link between the lensing potential and the parameter space. Reducing the surface area of boundary conditions in the parameter space can also reduce the number of local minima at the boundary surface. The fewer local minima there are the better one can find the global minimum. Priors on ($\theta$, $q$), i.e. based on the observed light distribution, must be transformed into priors on $(e_1, e_2)$ accordingly. In this work we assign uniform uninformative priors on $(e_1, e_2)$.

In general, the particular choice of the parameterization can be crucial. The smoother a change in parameter space reflects a small change in the model output, the better a convergence algorithm can deal with the system. The fewer constraints and boundary surfaces there are in the parameter space, the more general convergence algorithm manage to converge to the global minimum.

\subsubsection{Convergence with additional constraints}
In cases where the source galaxy hosts a quasar that dominates the luminosity, its lensed positions in the image plane can be determined by the data without knowledge of the lens, source position or the extended surface brightness. The feature in the image is very well predicted by the PSF model and dominates the brightness over an extended area in the image. Any proposed lens model that predicts the image positions displaced from the features in the image will be excluded by the data with high significance. The quasar point sources introduce a degeneracy of acceptable solutions within the original parameter space. Knowing about this degeneracy can lead to faster convergence.

When having $N$ bright point source images, there are $2N$ constraints to the system (their positions in the image plane). This reduces the effective dimensionality of the parameter space by $2N$. Lensing has three symmetries imprinted in the positional information: Two translations and one rotation. These transformations do not change the lens model apart from its own transformation.

In general, we can use any parameterization $\theta_i$ of an originally $M$-dimensional parameter space to dimension $n=M-2N+3$ (with $N>=2$) if there exists a bijective transformation (an exact one-to-one mapping of the two sets) to the original parameter space with the applied constraints. In the case of four bright images of a quasar, we determine an $(M-5)$-dimensional parameter space and solve for the source plane position of the quasar and five additional lens model parameter with a non-linear solver. This reduces the non-linear parameter space in the PSO process and leads to faster convergence without breaking any degeneracies. The choice of the five lens model parameters is arbitrary as long as the parameters can provide a solution ot the point source mapping. Priors on these parameters have also to be applied in the sampling process.

\subsubsection{Likelihood computation} \label{sec:likelihood_compute}
The likelihood calculation on the image level is the product of the likelihoods of each pixel \citep[see e.g.][for a similar approach]{Suyu:2013p4952}. We estimate the variance on the intensity at pixel $i$ as

\begin{equation}
	\sigma_{\text{pixel,i}}^2 = \sigma_{\text{bkgd,i}}^2 + fd_{\text{model,i}}
\end{equation}
where $\sigma_{\text{bkgd,i}}$ is the background noise estimated form the image, $d_{\text{model,i}}$ the model prediction at pixel $i$ and $f$ a scaling factor. A pure Poisson noise results in $f$ being the product of exposure time and gain. The likelihood of the data $d_{\text{data}}$ with $N_\text{d}$ image pixels given a model $d_{\text{model}}$ with non-linear lens model parameters ${\boldsymbol{\theta}}$ can then be written as a marginalization over the linear parameters $\boldsymbol{\xi}$, the source surface brightness parameters:
\begin{equation} \label{eqn:prob_formula}
	P(d_{\text{data}}|\boldsymbol{\theta}) = \int d\boldsymbol{\xi} P(d_{\text{data}}|\boldsymbol{\theta}, \boldsymbol{\xi}) P(\boldsymbol{\xi})
\end{equation}
where
\begin{equation} \label{eqn:pix_compare}
	P(d_{\text{data}}|\boldsymbol{\theta}, \boldsymbol{\xi}) = \frac{1}{Z_\text{d}} \exp \sum_{i=1}^{N_\text{d}} \left[ -\frac{(d_{\text{data,i}} - d_{\text{model,i}})^2}{2\sigma_{\text{pixel,i}}^2} \right].
\end{equation}
with $ \boldsymbol{d}_{\text{model}} = \boldsymbol{X}(\boldsymbol{\theta}) \boldsymbol{{\xi}}$.
$Z_\text{d}$ is the normalization
\begin{equation}
	Z_\text{d} = (2\pi)^{N_\text{d}/2} \prod_i^{N_\text{d}} \sigma_{\text{pixel,i}}
\end{equation}
and $P(\boldsymbol{\xi})$ the prior distribution of the shapelet coefficients. We assume a uniform prior distribution which is independent of the lens model. The integral in equation (\ref{eqn:prob_formula}) can be computed around the maximum $\boldsymbol{{\xi}_0}$ coming from equation (\ref{eqn:beta_max}) with covariance matrix $M^{\xi}$ from equation (\ref{eqn:cov_matrix}). With a second order Taylor expansion around $\boldsymbol{{\xi}_0}$, equation (\ref{eqn:prob_formula}) can be written as
\begin{equation} \label{eqn:p_delta}
	P(d_{\text{data}}|\boldsymbol{\theta}, \boldsymbol{{\xi}_0} + \Delta \boldsymbol{\xi}) \approx P(d_{\text{data}}|\boldsymbol{\theta}, \boldsymbol{{\xi}_0}) \cdot e^{-\frac{1}{2}\boldsymbol{\Delta\xi}^{T}(M^{\xi})^{-1}\boldsymbol{\Delta\xi}}.
\end{equation}
Integrating equation (\ref{eqn:p_delta}) over $\Delta \boldsymbol{\xi}$ results in 
\begin{equation}
	P(d_{\text{data}}|\boldsymbol{\theta}) = P(d_{\text{data}}|\boldsymbol{\theta}, \boldsymbol{{\xi}_0}) \left[(2\pi)^m\text{det}(M^{\beta})\right]^{\frac{1}{2}}
\end{equation}

In principle, equation (\ref{eqn:pix_compare}) is the cost function to use for image comparison. The information about the image positions is included in this cost function. The problem with this cost function is that convergence to a good model can be difficult. The use of additional or derived information, such as the explicit image positions, can facilitate convergence.

\subsubsection{Steps towards convergence}
\begin{figure*}
  \centering
  \includegraphics[angle=0, width=120mm]{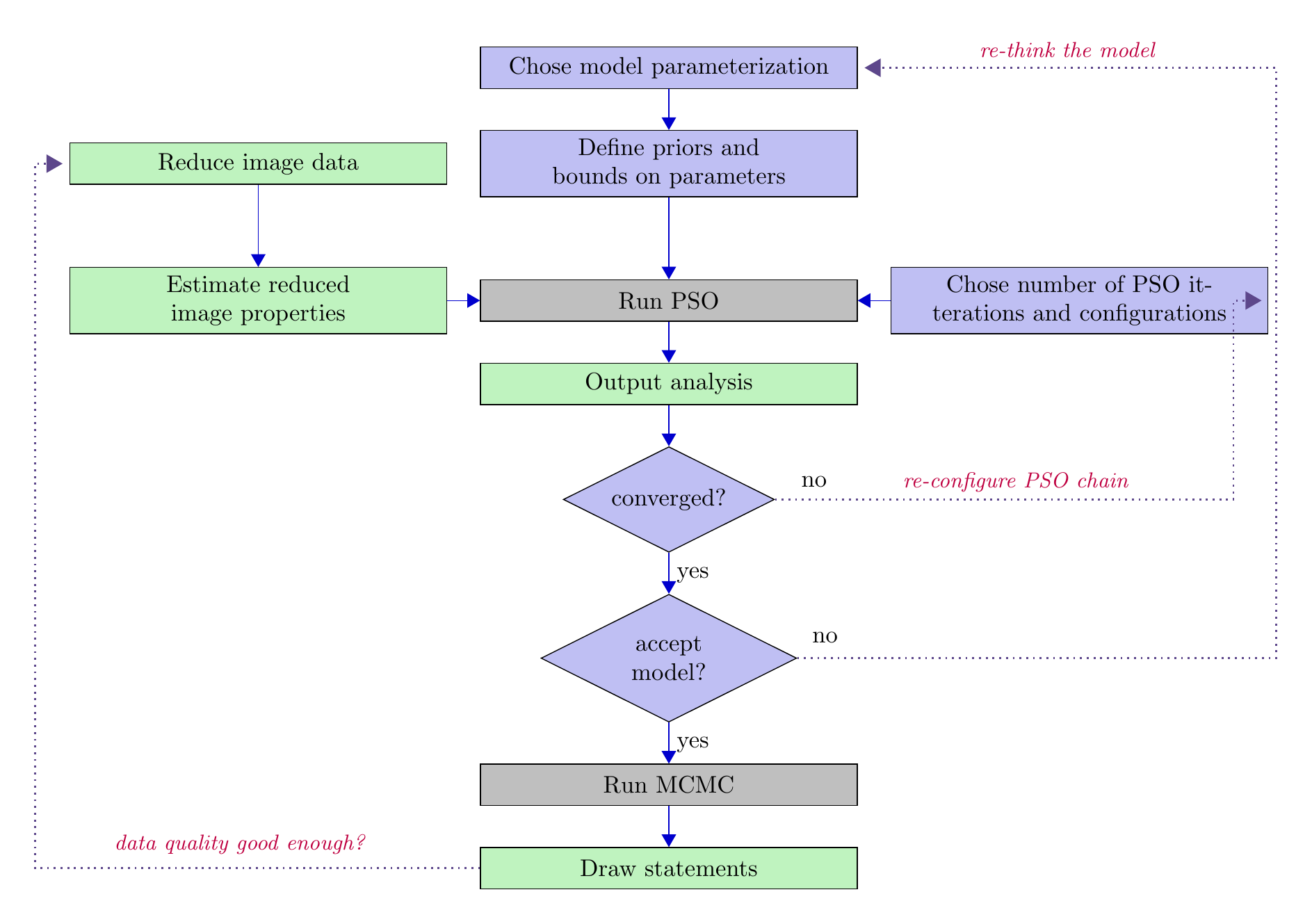}
  \caption{Chart of the framework highlighting user interactions. Human interactions are needed for some tasks (green) and decisions (blue). Automated tasks are shown in gray. The core of this framework is to clearly split the preprocessing from the fitting algorithm.}
\label{fig:table_model}
\end{figure*}

Having presented our model parameterization in Section \ref{sec:basis_sets} and discussed certain aspects of model fitting and convergence in the previous paragraphs, we describe our steps which allows us to find a reasonable fit to the data. Figure \ref{fig:table_model} illustrates the framework. Prior to the convergence algorithm, the image data has to be analyzed, the model configuration has to be chosen, the prior values have to be set and the specific configuration of the PSO process have to be given as an input. The fitting should be done within an automated process where no interaction of the modeler is needed. The output of the PSO run can then be analyzed by the modeler in terms of convergence and quality of fitting. This may lead to a change in the model parameters, functions, configuration etc and the process is run again. Once convergence is achieved and the fitting result is good, the MCMC process is run to map out the valid parameter space given the model parameters chosen. Figure \ref{fig:mock_triangle} illustrates the PSO and MCMC process in a 9 dimensional parameter space. The thin blue lines corresponds to the PSO process. Once this process is converged we start the MCMC process around this position (light and dark gray contours). Certain parameters are more degenerate than others. We try to map the parameter space such that remaining degeneracies are controllable.

\section{Example - RXJ1131-1231} \label{sec:example} 
In the following, we test our method on the gravitational lens RXJ1131-1231. This lens was discovered by \cite{Sluse:2003p8680} and the redshift of the lens $z_l = 0.295$ and of the background quasar source $z_s = 0.658$ was determined spectroscopically. The lens was extensively modeled by \cite{Claeskens:2006p8872, Brewer:2008p8808, Suyu:2013p4952}.
We use the same archival HST ACS image in filter F814W (GO 9744; PI: Kochanek) as \cite{Suyu:2013p4952} for our lens modeling and follow a similar procedure for the reduction process and error estimation. We make use of the \texttt{MultiDrizzle} product from the HST archive. The PSF is estimated from stacking of nearby stars. We estimate a PSF model error by computing the variance in each pixel from the different stars after a sub-pixel alignment with an interpolation done using all the stars. We assume that this model error is proportional to the intensity of the point source. This method is meant to demonstrate our method in fitting the best configuration. The lens model is parameterized as a SPEP (ellipsoid) and a second SPP (round) profile (see Eqn \ref{eqn:SPEP}). Furthermore we choose 15 shapelet basis sets in the potential and a constant external shear component. For the lens light we follow \cite{Suyu:2013p4952} and use two elliptical Sersic profiles with common centroids and position angles to describe the main lens galaxy and a circular S\'ersic profile with $n_{\text{sersic}} = 1$ for the small companion galaxy.

\begin{figure*}
  \centering
  \includegraphics[angle=0, width=180mm]{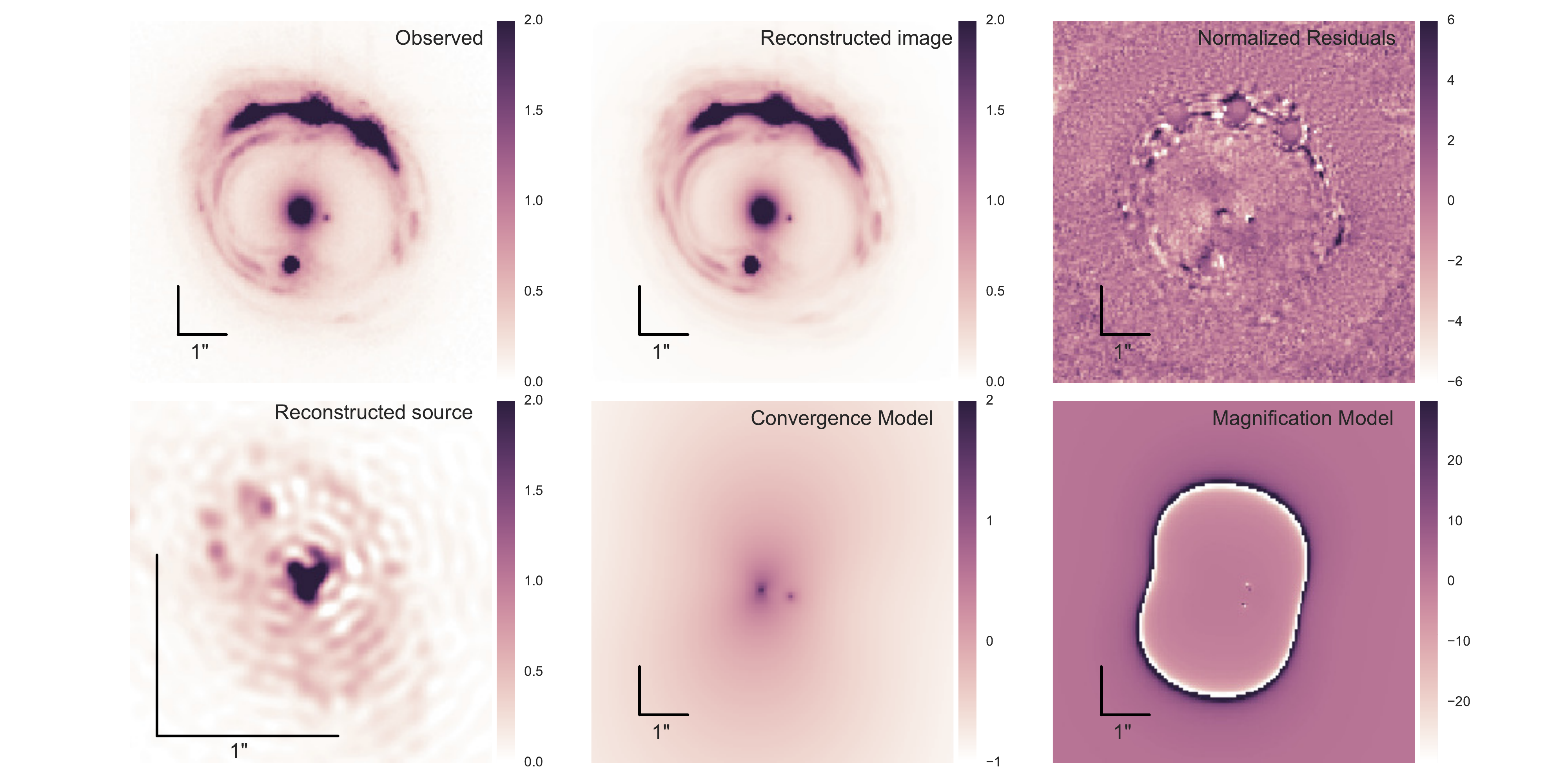}
  \caption{Modeling of RXJ1131-1231 HST ACS F814W image. Upper left: Observed image. Upper middle: Best fit pre-construction. Upper right: normalised residuals of the reconstruction. Lower left: Reconstructed source with 1326 shapelet coefficients (up to order 50). Lower middle: Convergence model of the lens. Lower right: Magnification model of the lens.}
\label{fig:RXJ1131}
\end{figure*}

Figure \ref{fig:RXJ1131} shows our result of the fitting process to the HST image. In the upper left panel we show the reduced data. Upper middle shows the best fit model. On the upper right the normalized residuals are plotted. The reduced $\chi^2$ value of this fit is $\chi^2_{\text{red}} = 1.5$ without adjusting any Poisson factors nor the background noise level originally derived from the image data products. We clearly see that there are significant residuals around the point sources which indicates clearly that our PSF model needs further improvement and that even our error model on the PSF seems to underestimate the model error in certain regions. Furthermore, extended regions of over- or under-fitting indicate that the lens model can be improved. Source surface brightness adoptions could have acted to reduce the error in the fit in case of a perfect lens model. The lower left panel shows the reconstructed extended source surface brightness profile. We clearly see the presumably star forming clumps which lead to the features in the extended Einstein ring. In the lower middle panel of Figure \ref{fig:RXJ1131} our lens model is shown in terms of the convergence map. We notice that without mass-to-light priors, the position of the two modeled clumps is strikingly close to the position of the luminous galaxy and its companion. In the lower right panel, the magnification map is shown. The reconstruction of the image depends on the number of shapelet coefficients used. In Appendix \ref{app:n_max} we discuss the effect of $n_{max}$ on the quality of the source and image reconstruction for this particular lens system.

Comparing different lens and source model reconstructions from different methods is difficult. Different source surface brightness reconstruction techniques use different number of parameters and thus can have different $\chi^2$ values without changing the lens model. Error models and masking do have a significant impact on $\chi^2$. Setting priors may also lead to different results (In case of \cite{Suyu:2013p4952} the position of the sub-clump modeled as a singular isothermal sphere was fixed at the position of the luminous companion and additional information from velocity dispersion measurements). All in all, different lens modeling techniques can only be properly compared based on mock data. And even on mock data, different input types of lens and source models might have a significant influence on the relative performance of the methods.

\section{Detectability of substructure}\label{sec:detection}
One of our main focuses for the model we present is to find and quantify substructure within a lens. In this section, we want to discuss the following issues:
\begin{enumerate}
	\item To what extend are our model basis functions and description able to reproduce the true image? \label{Q1}
	\item In case of a perfect modeling: Are we able to recover the true parameter configuration in a large parameter space? \label{Q2}
	\item In case of an imperfect modeling: How does this affect the sensitivity limit, finding and quantification of substructures? \label{Q3}
\end{enumerate}

To answer our first question, we refer to our data example of RXJ1131-1231 in Section \ref{sec:example} of Figure \ref{fig:RXJ1131}. Even though the observed and predicted images can hardly be distinguish by eye, the residual map indicates room for improvements in our modeling. Nevertheless the fact that our mass-to-light prior-free lens model provides us with a realistic solution might indicate that we are not far from reality. A priori, we do not know whether the solution found in Section \ref{sec:example} is the global minimum of the parameter space chosen and therefore the best reachable solution within the choices and parameters made. We will investigate whether the finder algorithm is able to recover the true input parameters when fitting mock images in the next section.

\subsection{Substructure finding}\label{sec:detection_prob}
\begin{figure*}
  \centering
  \includegraphics[angle=0, width=160mm]{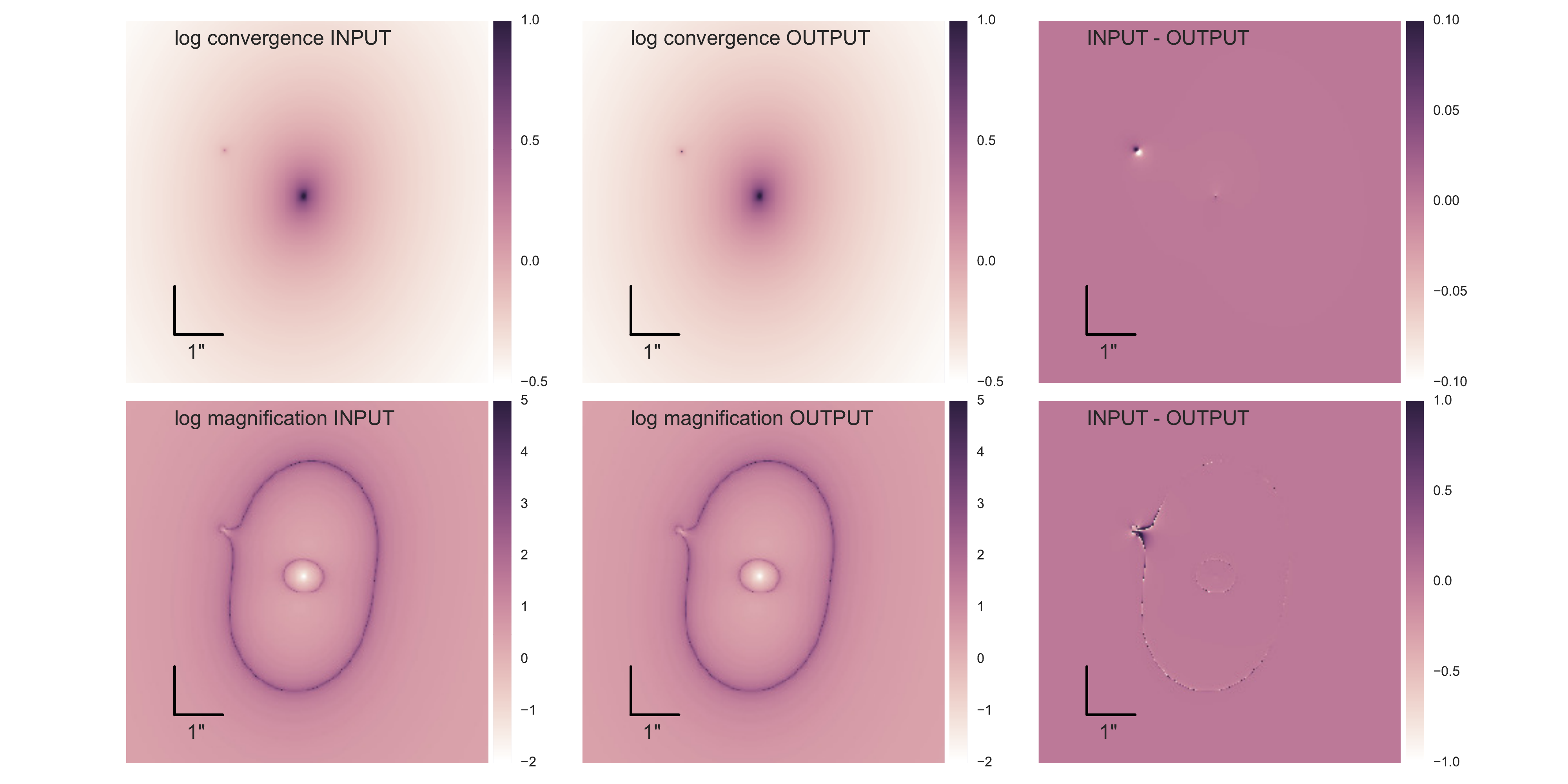}
  \caption{Lens model input-output comparison for convergence and magnification log(abs(magnification)). Without priors on the size or position of the sub-clump of order $10^{8}M_{\odot}$, the PSO can find the lens configuration. The input-output comparison of the image is illustrated in Figure \ref{fig:reconstruction}. The sensitivity for the sub-clump detection is analyzed in Figure \ref{fig:mock_triangle}. We clearly see that the we are sensitive to the position up to 0.1".}
\label{fig:subclump_reconstruction}
\end{figure*}

To approach question \ref{Q2} above we take a mock image that is highly inspired by RXJ1131-1231 of section \ref{sec:example}. We keep the image quality fixed (i.e. noise levels, pixel size and PSF) but change the lens model such that we have one big SPEP profile and a minor sub-clump, a spherical power law potential (SPP).
Ideally, we do not want to set any prior on the position, mass, shape and number of substructures. If we were interested in luminous sub-structure we could add mass-to-light priors. As we want to use our method to potentially detect dark sub-structure, we are not allowed to give any mass-to-light prior. Therefore we want to check whether our algorithm finds the preferential parameter space in the model. The main focus is on the position of the sub-clump. To explore our capability of finding sub-clumps, we generate mock data with a sub-clump in the lens model at a random position. We add Poisson and Gaussian noise on the mock image. We then run the convergence method on that image with the same weak prior information as was done for the real image in section \ref{sec:example}. We repeat this procedure 10 times. Our result is:
\begin{itemize}
	\item Success rate in position of 100\%. For our setting with a random sampling of the prior parameter space, all the runs ended around the right solution (PSO).
	\item Detectability down to $10^{-4}$ level of the total lens mass in the arc of the Einstein ring (MCMC).
	\item Time for convergence of about $10^5$ evaluations of a model configuration needed. One evaluation takes few seconds.
\end{itemize}
For one realization of the input-output process the comparison is shown in Figure \ref{fig:subclump_reconstruction} in terms of convergence and magnification and their residuals. We clearly see that the position of the sub-clump can be well recovered and the appearance of the critical line do match very well. This means that there is no other degenerate solution within the parameter space that can reproduce a similar feature like a sub-clump no matter what combination of source surface profile and lens model we chose.
This test shows that with a ideal model we can find a single sub-clump in the lens mass without any prior on its existence and position. We also highlight that the relative likelihood comparison is large (more than 5 sigma compared with the best fit model without a sub-clump). This statement in this form can only be made if other effects (such as error in the PSF model) do not interfere. As we showed for the real image in section \ref{sec:example}, errors in the PSF model alone can potentially lead to a higher increase in the minimal $\chi ^2$. This test together with the finding of a sub-clump in RXJ1131-1231, in both cases without setting priors on position, mass-to-light, concentration and mass, are encouraging hints that our model approach can extract valuable information about a lens system in a rather unbiased way. 

\subsection{Substructure sensitivity}\label{sec:detection_limit}
The last section discussed the potential to recover sub-structure. We showed that this is possible for substructure with mass ratios of $M_{\text{sub}} \sim 10^{-4}M_{\text{lens}}$ consistent with a direct detection of a sub-clump in \cite{Vegetti:2012p4937}. This implies, that we are sensitive to this mass regime in fitting a mass-concentration relation. The concentration of the sub-clump is not exactly matched when comparing with the most likely solution of the PSO in Figure \ref{fig:subclump_reconstruction}. To effectively see how well we can constrain certain parameters in the model from the data, we do a full likelihood analysis with a MCMC. The mapping of the entire parameter space of this realization with an MCMC is illustrated in Figure \ref{fig:mock_triangle} for exactly the same realization as Figure \ref{fig:subclump_reconstruction}. The red lines mark the input parameters. We see, that the input parameters are always within 2 sigma of the output parameter distribution. We see that there is a partial degeneracy between mass and concentration of the sub-clump ($\gamma$ (clump) and log $\theta_E$ (clump) in Figure \ref{fig:mock_triangle}. Not surprising, it is difficult to constrain the profile of a very small clump which itself is close to the detection limit. Even better data than HST, such as JWST or ALMA can potentially detect clumps down to lower mass levels and also constrain the profiles of these small clumps. The sensitivity limit relies mostly on three criteria: FWHM of PSF, magnification at the position of the sub-clump and source surface brightness variation at the lensed position of the sub-clump. For any different data qualities, telescopes etc, we are able to perform such sensitivity tests.

\section{Conclusion} \label{sec:conclusion}
In this paper, we introduced a new strong lensing modeling framework which is based on versatile basis sets for the surface brightness and lens model. We identified the following aspects of our framework:
\begin{itemize}
	\item Its modular design allows for a step-by-step increase in complexity. We are able to determine which part of the modeling needs more complexity to reproduce a lens system. 
	\item It allows for automated or semi-automated fitting procedures. An adaptive cost function combined with a best fit algorithm, allows it to fit different strong lens systems without giving specific priors to each one of them. This allows for faster and more systematic analyses of large numbers of lens systems.
	\item It is suitable for a wide range of strong lensing systems and observing conditions. Our framework can be applied to various levels of image quality, different type of lens system, sizes of the lens, etc, as the convergence algorithm does not rely on strong initial priors.
	\item It features fast source reconstruction techniques. The evaluation of the cost function given a position in parameter space including the simulation of the image can be achieved within seconds. Furthermore our convergence algorithm allows for massive parallelization on a distributed computer architecture.

\end{itemize}
We further proposed a way to model strong lens systems to extract information about the substructure content within the lens. Such investigations can potentially provide useful constraints on the abundances of low mass objects. To learn about the dark matter properties from strong lensing, one needs to combine well chosen descriptions for the source light and lens mass, algorithm techniques which can find solution in high dimension parameter spaces and a combination of different data sets to break degeneracies (multi-band imaging, spectroscopy, etc.). A special focus has to be made in choosing the right set of basis functions and the algorithmic design of the convergence method. Our approach is encouraging as it succeeds in recovering substructure in the lens without setting mass-to-light priors.

\begin{acknowledgements}
SB thanks Sebastian Seehars, Claudio Bruderer, Andrina Nicola, Frederic Courbin, Sherry Suyu and Kevin Fusshoeller for helpful comments and useful discussions and Joel Akeret for valuable inputs on the software design and algorithms. We also want to thank the anonymous Referee who helped improving this manuscript. We acknowledge the import, partial use or inspiration of the following python packages: CosmoHammer \citep[][]{Akeret:2013p8317}, Ufig \citep[][]{Berge:2013p5329}, Hope \citep[][]{Akeret:2015p8687}, astropy http://www.astropy.org/, triangle , numpy, scipy. This work has been supported by the Swiss National Science Foundation (grant number 200021\_143906 and 200021\_14944).
\end{acknowledgements}

\bibliography{papers_bibtex}{}

\appendix

\section{Shapelets} \label{app:shapelets}
\begin{figure*}
  \centering
  \includegraphics[angle=0, width=120mm]{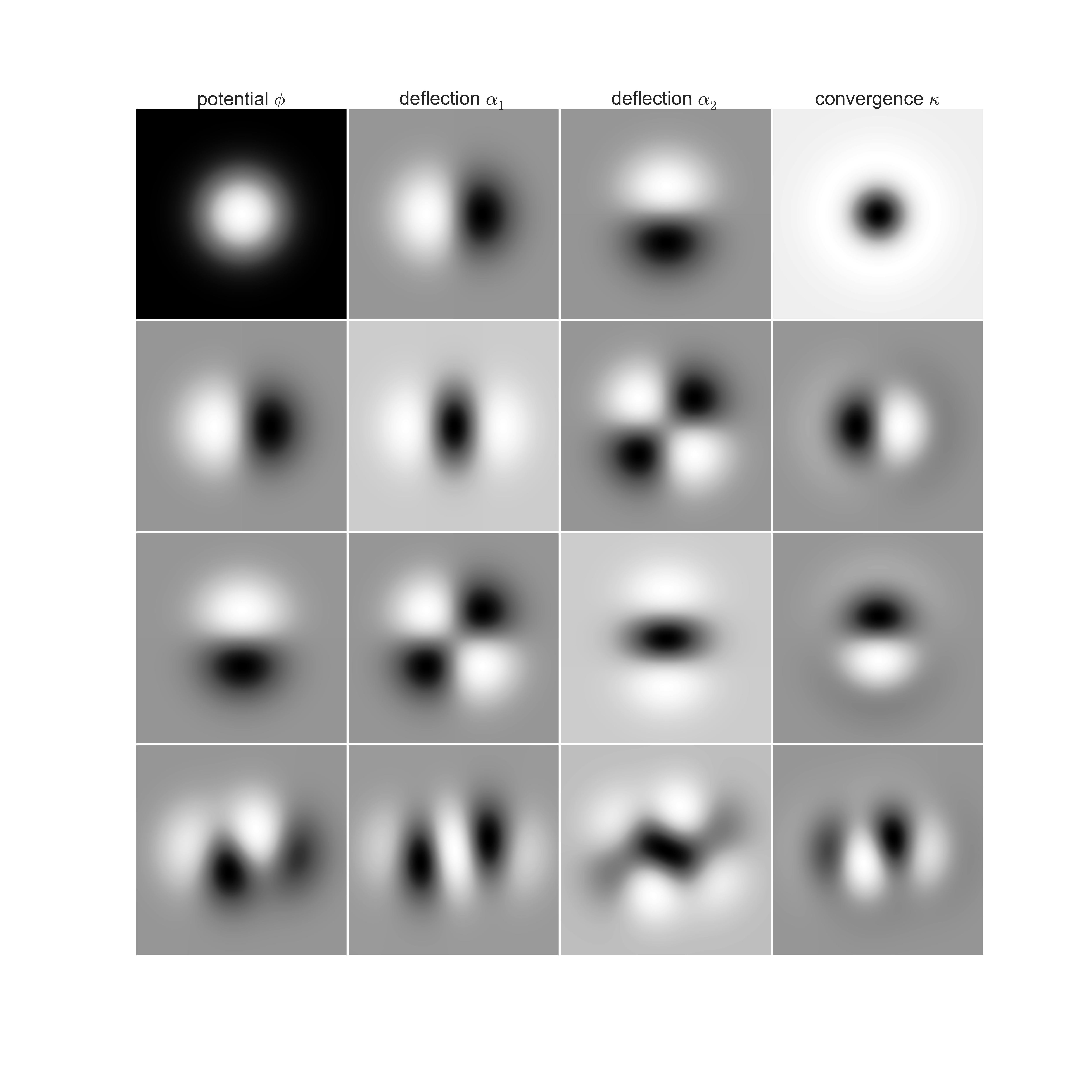}
  \caption{The shapelet functions in potential space are plotted in the first row. From top to bottom: (0,0), (1,0), (0,1), (0,1) + (3,0). The second and third rows show the deflection angles $\alpha_1$ and $\alpha_2$. The last row shows the corresponding convergence $\kappa$.}
\label{fig:shapelets_potential}
\end{figure*}

The two dimensional Cartesian shapelets as described in \cite{Refregier:2003p8153} are the multiplication of two one-dimensional shapelets $\phi(x)$:
\begin{equation}
	\phi_{\textbf{n}}(\textbf{x}) \equiv \phi_{n_1}(x_1) \phi_{n_2}(x_2).
\end{equation}
The one-dimensional Cartesian shapelet is given by:
\begin{equation}
	\phi_n(x) \equiv \left[2^n\pi^{\frac{1}{2}}n! \right]^{-\frac{1}{2}} H_n(x) e^{-\frac{x^2}{2}}
\end{equation}
where $n$ is a non-negative integer and $H_n$ the Hermite polynomial of order $n$. The dimensional basis function is described as
\begin{equation}
	\phi_n(x;\beta) \equiv \beta^{-\frac{1}{2}} \phi(\beta^{-1}x).
\end{equation}
In two dimensions, we write
\begin{equation}
	\bf{B}_{\bf{n}}(x;\beta) \equiv \beta^{-1} \phi_{\bf{n}}(\beta^{-1}\bf{x}).
\end{equation}
This set of basis functions is used for the source surface brightness modeling and the lensing potential. To find the derivatives of this functions, \cite{Refregier:2003p8153} introduced raising and lowering operators which act on the basis functions as
\begin{equation}
\hat{a}\phi_n = \sqrt{n}\phi_{n-1}, \hat{a}^\dagger \phi_n = \sqrt{n+1}\phi_{n+1},
\end{equation}
the derivative operator can be written as
\begin{equation}
	\frac{d}{dx} = \frac{1}{\sqrt{2}}\left(\hat{a} - \hat{a}^\dagger \right)
\end{equation}
and therefore any derivative can be written as a superposition of two other shapelet basis functions (for further discussions, see \cite{Refregier:2003p8153}) In Figure \ref{fig:shapelets_potential}, it is illustrated how shapelet basis functions in the potential space do map in the deflection angle and convergence.

\section{Number of shapelet coefficients} \label{app:n_max}
\begin{figure*}
  \centering
  \includegraphics[angle=0, width=180mm]{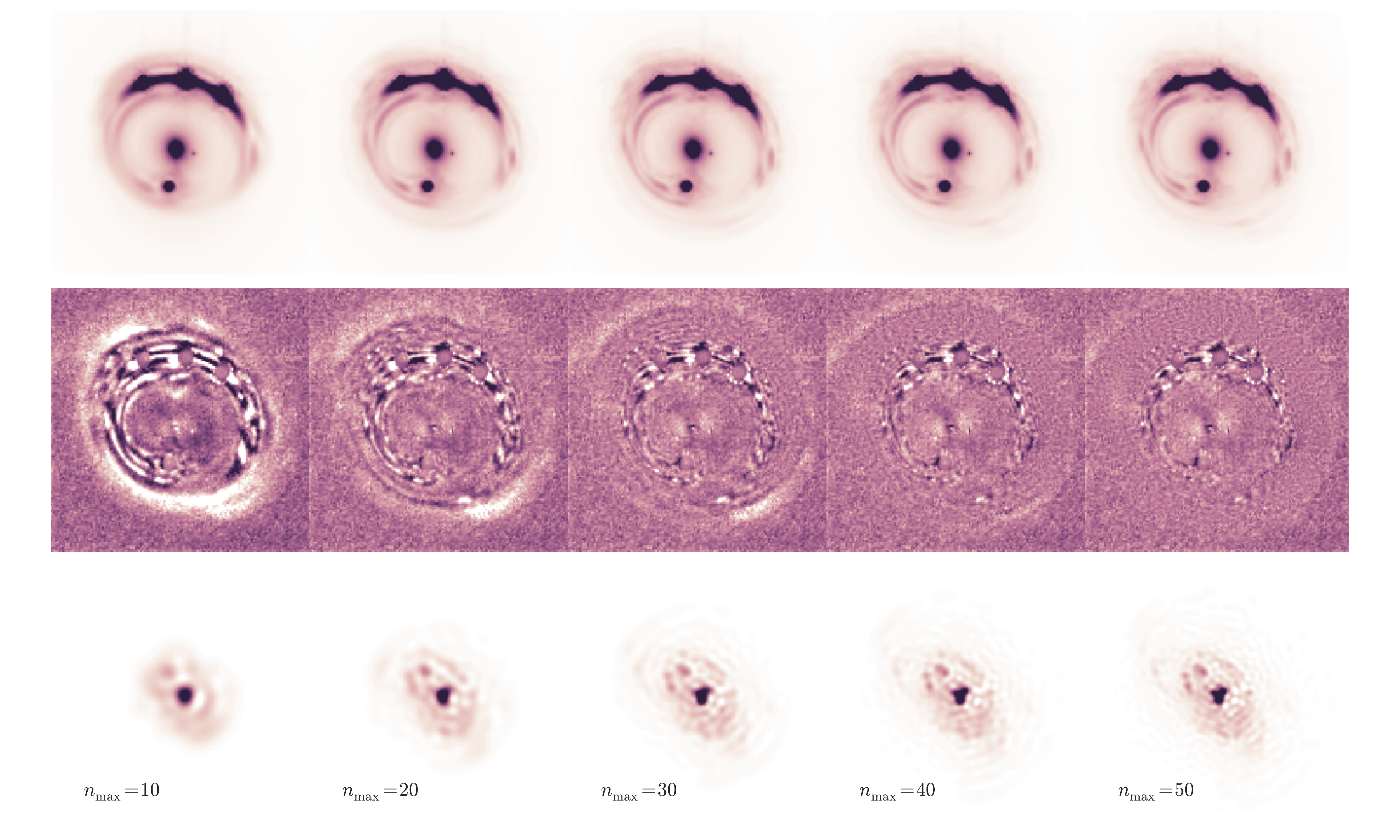}
  \caption{The source surface brightness reconstruction of the lens system RXJ1131-1231 is modeled with different shapelet orders $n_{\max}$. Upper panel: The reconstructed image. Middle panel: The normalized residual maps. Lower panel: The reconstructed source. From left to right: Increasing number of shapelet order $n_{\max}$ from $10$ to $50$.}
\label{fig:num_shapelet}
\end{figure*}

The choice of the maximal order of the shapelet coefficients $n_{\max}$ and its corresponding number $m = (n_{\text{max}} + 1) \cdot (n_{\text{max}} +2)/2$ has a significant influence in the goodness of fit to imaging data. In Figure \ref{fig:num_shapelet} we illustrate this by reconstructing the source surface brightness with different $n_{\max}$. We use the same lens model as in Figure \ref{fig:RXJ1131}. We see that even with $n_{\max} = 10$, most of the features in the arcs of the image could be reconstructed qualitatively but significant residuals remain. By increasing $n_{\max}$, more and more details in the source appear and the residuals go down.

\section{Particle Swarm Optimization} \label{app:pso}
\begin{figure*}
  \centering
  \includegraphics[angle=0, width=180mm]{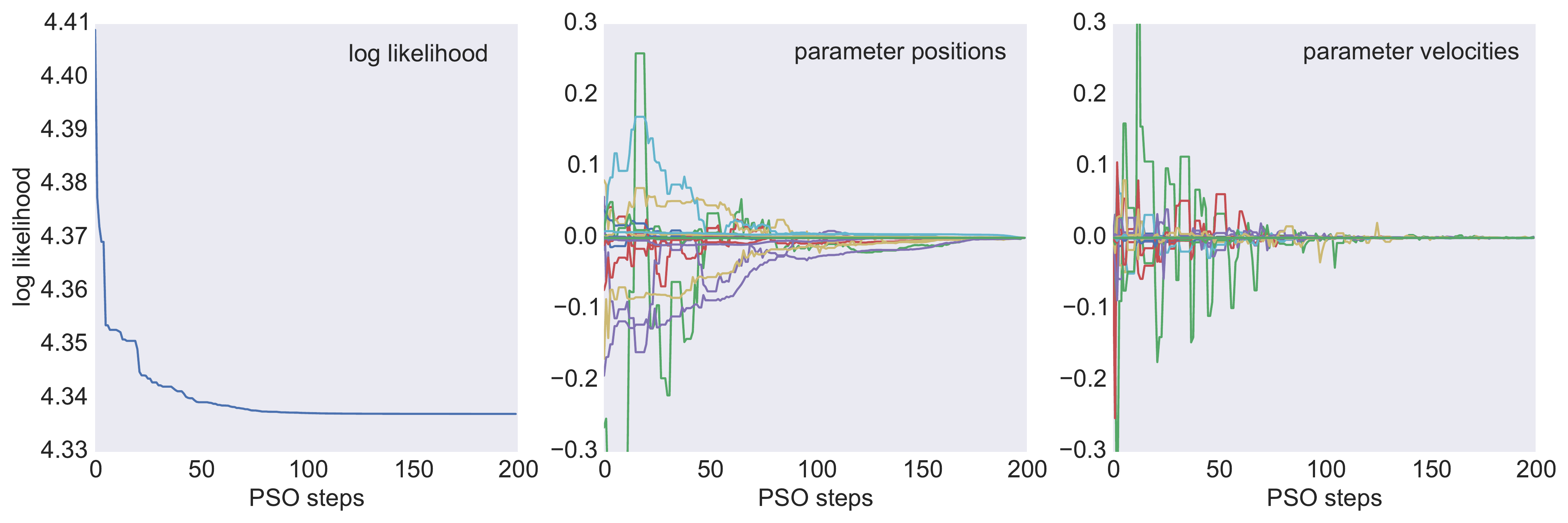}
  \caption{Illustration of the PSO process in 20 dimensions with 160 particles and 200 iterations. Left panel: Evolution of the log likelihood of the best fit particle. Middle panel: The difference of the parameter values from the best fit at each iteration relative to the end point of the PSO process. Right panel: Velocity of the best fit particle at each iteration. Different colors are used for each of the parameters.
  }
\label{fig:pso_walking}
\end{figure*}

The Particle Swarm Optimization (PSO) description was introduced by \cite{Kennedy:2001p8447} as a method to find the global minima in a high dimensional non-linear distribution. The algorithm is motivated by the physical picture of a swarm of particles moving in a physical potential. Every particle gets assigned a position in parameter space, a function evaluation (the log likelihood value) and a velocity in parameter space. The particles is assigned a swarm "physical" behavior when moving up or downwards a potential and a "swarm" behavior when redirecting their velocity towards the particle at the deepest place of the potential. The PSO process is illustrated in Figure \ref{fig:pso_walking} in a 20 dimensional parameter space.
The implementation of the PSO algorithm used in this work is publicly available as part of the {\tt CosmoHammer} \cite{Akeret:2013p8317} software package. The inertia weight strategy comes from \cite{Bansal:2011p8496} and the stopping criteria of \cite{Zielinski:2008p8432} was implemented.

\end{document}